\documentclass[8.5pt,twoside,twocolumn]{article}
\usepackage[super,sort&compress,comma]{natbib}
\usepackage{mhchem}
\usepackage{bm}
\usepackage{times,mathptmx}
\usepackage{graphicx}
\usepackage{amsmath}
\usepackage{amssymb}
\usepackage{pifont}

\usepackage{epstopdf}

\usepackage{sectsty}
\usepackage{balance}
\usepackage{graphicx} 
\usepackage{lastpage}
\usepackage[format=plain,justification=raggedright,singlelinecheck=false,font=small,labelfont=bf,labelsep=space]{caption}
\usepackage{fancyhdr}

\newcommand{\Ub}{\mathbf{U}}
\def\nablab{\mbox{\boldmath$\nabla$\unboldmath}}
\newcommand{\kb}{\mathbf{k}}
\def\mean#1{\langle #1 \rangle}

\begin{document}

\thispagestyle{plain}
\fancypagestyle{plain}{
\fancyhead[L]{
}
\fancyhead[C]{\hspace{-1cm}
}
\fancyhead[R]{
\vspace{-0.2cm}}
\renewcommand{\headrulewidth}{1pt}}
\renewcommand{\thefootnote}{\fnsymbol{footnote}}
\renewcommand\footnoterule{\vspace*{1pt}%
\hrule width 3.4in height 0.4pt \vspace*{5pt}}
\setcounter{secnumdepth}{5}

\makeatletter
\def\subsubsection{\@startsection{subsubsection}{3}{10pt}{-1.25ex plus -1ex minus -.1ex}{0ex plus 0ex}{\normalsize\bf}}
\def\paragraph{\@startsection{paragraph}{4}{10pt}{-1.25ex plus -1ex minus -.1ex}{0ex plus 0ex}{\normalsize\textit}}
\renewcommand\@biblabel[1]{#1}
\renewcommand\@makefntext[1]%
{\noindent\makebox[0pt][r]{\@thefnmark\,}#1}
\makeatother
\renewcommand{\figurename}{\small{Fig.}~}
\sectionfont{\large}
\subsectionfont{\normalsize}

\fancyfoot{}
\fancyfoot[LO,RE]{\vspace{-7pt}
}
\fancyfoot[CO]{\vspace{-7.2pt}\hspace{12.2cm}
}
\fancyfoot[CE]{\vspace{-7.5pt}\hspace{-13.5cm}
}
\fancyfoot[RO]{\footnotesize{\sffamily{1--\pageref{LastPage} ~\textbar  \hspace{2pt}\thepage}}}
\fancyfoot[LE]{\footnotesize{\sffamily{\thepage~\textbar\hspace{3.45cm} 1--\pageref{LastPage}}}}
\fancyhead{}
\renewcommand{\headrulewidth}{1pt}
\renewcommand{\footrulewidth}{1pt}
\setlength{\arrayrulewidth}{1pt}
\setlength{\columnsep}{6.5mm}
\setlength\bibsep{1pt}

\twocolumn[
  \begin{@twocolumnfalse}
\noindent\LARGE{\textbf{Flow in channels with superhydrophobic trapezoidal textures }}
\vspace{0.6cm}

\noindent\large{\textbf{Tatiana V. Nizkaya,\textit{$^{a}$} Evgeny S. Asmolov,\textit{$^{a,b,c}$} Olga I. Vinogradova\textit{$^{a,d,e,\ast}$} }\vspace{0.5cm}}

\noindent\textit{\small{\textbf{Received Xth XXXXXXXXXX 20XX, Accepted Xth XXXXXXXXX 20XX\newline
First published on the web Xth XXXXXXXXXX 200X}}}

\noindent \textbf{\small{DOI: 10.1039/b000000x}}
\vspace{0.6cm}

\noindent \normalsize{  Superhydrophobic one-dimensional surfaces reduce drag and generate transverse hydrodynamic phenomena by combining hydrophobicity and roughness to trap gas bubbles in microscopic textures. Recent works in this area have focused on specific cases of superhydrophobic stripes. Here we provide some theoretical results to guide the optimization of the forward  flow and transverse hydrodynamic phenomena in a parallel-plate channel with a superhydrophobic trapezoidal texture, varying on scales larger than the channel thickness. The permeability of such a thin channel is shown to be equivalent to that of a striped one with greater average slip. The maximization of a transverse flow normally requires largest possible slip at the gas areas, similarly to a striped channel. However, in case of trapezoidal textures with a very small fraction of the solid phase this maximum occurs at a finite slip at the gas areas. Exact numerical calculations show that our analysis, based on a lubrication theory, can also be applied for a larger gap. However, in this case it overestimates a permeability of the channel, and underestimates an anisotropy of the flow. Our results provide a framework
for the rational design of superhydrophobic surfaces for microfluidic applications.

 }\vspace{0.5cm}
 \end{@twocolumnfalse}
]

\footnotetext{\textit{$^{a}$~A.N.~Frumkin Institute of Physical Chemistry and
Electrochemistry, Russian Academy of Sciences, 31 Leninsky
Prospect, 119071 Moscow, Russia}}
\footnotetext{\textit{$^{b}$~Central Aero-Hydrodynamic Institute, 140180
Zhukovsky, Moscow region,  Russia}}
\footnotetext{\textit{$^{c}$~Institute of Mechanics, M.V.~Lomonosov Moscow State University, 119991 Moscow,
Russia}}
\footnotetext{\textit{$^{d}$~Faculty of Physics, M.V.~Lomonosov Moscow State University, 119991 Moscow, Russia; E-mail: oivinograd@yahoo.com }}
\footnotetext{\textit{$^{e}$~DWI,  RWTH Aachen, Forckenbeckstr. 50, 52056 Aachen, Germany}}
\newcommand{\p}{\partial}
\newcommand{\eps}{\varepsilon}
\date{\today}

\section{Introduction}

Superhydrophobic (SH) textured materials have been extensively investigated for the last decade, since texture can bring exceptional wetting properties.~\cite{quere.d:2008,bico.j:2002} Beside that, SH materials in the Cassie
state, where the texture is filled with gas, are important in context of fluid dynamics and their superlubricating properties.~\cite{vinogradova.oi:2012,mchale.g:2010,rothstein.jp:2010} The drag reduction can be quantified by an effective slip length of the heterogeneous SH surface, which can be of the order of several $\mu$m or even more,~\cite{maali.a:2012,truesdell.r:2006,joseph.p:2006,tsai.p:2009} and for anisotropic surfaces varies with the orientation of the wall texture
relative to flow.~\cite{ou2005,tsai.p:2009}
Despite significant advances in the field, previous investigations of anisotropic textures were limited mostly by SH stripes. The problem of flow past striped SH
surfaces has previously been studied in the context of a reduction of
pressure-driven or shear forward flow in
thick~\cite{lauga.e:2003,belyaev.av:2010a,priezjev.nv:2005,ng:2009},
thin~\cite{feuillebois.f:2009} and intermediate \cite{TeoKhoo:09, ng:2010} channels, and it is directly relevant for
mixing,~\cite{feuillebois.f:2010b,vinogradova.oi:2011} and for a generation of a
tensorial electro-osmotic
flow.~\cite{bahga:2010,vinogradova.oi:2011,belyaev.av:2011a}

\begin{figure}[h]
\includegraphics[width=0.45\textwidth]{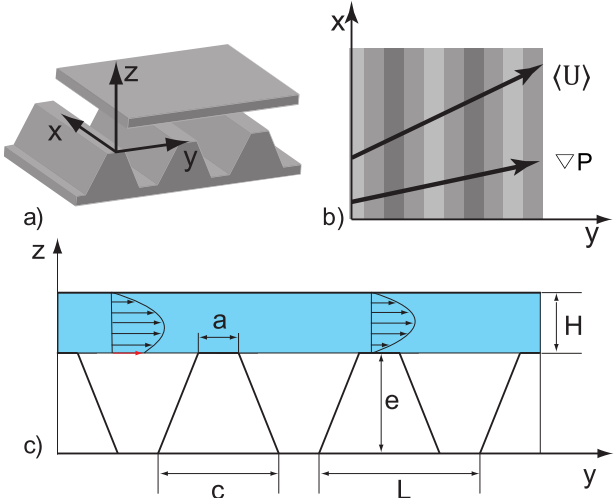}
\caption{ \it (a) Sketch of the channel with a hydrophilic (top) and a SH (bottom) wall with one-dimensional trapezoidal texture. (b) Illustration of tensorial hydrodynamic response,  with the direction of a flow misaligned from the pressure gradient (top view). (c) The cross-section on the channel flow with the trapezoidal texture.  }\label{sketch_trap}
\end{figure}

In this paper we focus on the trapezoid surface texture as illustrated in Fig.~\ref{sketch_trap}. This one-dimensional patterned SH surfaces, where the local (scalar) slip
length $b(y)$ varies only in one direction, is more general since it provides more geometrical texture parameters, namely, trapezoid base width, angle, height, and spacing. Furthermore, the trapezoid texture can be extended to its extreme, saw-tooth case, where the solid area vanishes. It can also be transformed to a striped texture. These grooved surfaces with a trapezoidal relief can be made easily with a modern soft lithography techniques, and were used in wetting studies.~\cite{herminghaus.s:2008} Note that such trapezoidal textures can be naturally formed from dilute rectangular grooves as a result of elasto-capillarity.~\cite{tanaka.t:1993,bico.j:2004,geim.ak:2003} They provide a stable Cassie state~\cite{li.w:2010} and can sustain very small areas of liquid/solid contact without losing mechanical stability. This type of surfaces has already been intensively used in the slippage experiment.~\cite{choi.ch:2006} However, how the trapezoidal relief of the texture impacts transport and hydrodynamic slippage remains largely unknown. Here we investigate a pressure-driven flow in such a channel, by focussing mostly on the so-called thin-channel limit, where a finite  thickness, $H$, is small compared to a texture period, $L$.

 Our paper is arranged as follows. In Section \ref{SecModel} we define the problem. The general expansions for the eigenvalues of
the permeability tensor of a thin channel, which fully describe its behavior at macrolevel, are constructed in Section \ref{SecPerm}. Then, in Section \ref{SecProp} we present specific results for a channel with a trapezoidal textured wall. Here beside analytical results obtained in a thin- channel limit, we present exact numerical data obtained at arbitrary channel gaps.  We conclude in Section \ref{SecConc}. In Appendix~\ref{local} we give some simple arguments suggesting that in case of shallow grooves the local slip profile follows the relief of the texture. Appendix~\ref{numeric} describes details of our numerical methods.

\section{Model\label{SecModel}}

Here we present the basic assumptions of our theoretical model, and define local slip boundary conditions.

We consider a pressure-driven flow in a channel consisting of two parallel walls located at $z=0$ and $z=h$ and
unbounded in the $x$ and $y$ directions as sketched in
Fig.\ref{sketch_trap}. The upper plate represents a no-slip hydrophilic
surface, and the lower plate is a SH surface. The origin of coordinates is
placed at the center of the solid
sector. The $x$-axis is directed along the texture, $z$-axis is orthogonal to the channel walls and the $y$-axis is normal both to $x$ and $z$. This (SH vs.
hydrophilic) geometry is relevant for various setups, where
the alignment of opposite textures is inconvenient or difficult. Besides,
the advantage of such a geometry is that it allows to avoid the gas bridging
and long-range attractive capillary forces,~\cite{andrienko.d:2004} which
appear when we deal with interactions of two hydrophobic
solids.~\cite{yakubov:00,tyrrell.jwg:2001}

As in most previous
publications,~\cite{bocquet2007,belyaev.av:2010b,priezjev.nv:2005,lauga.e:2003,maynes2007,asmolov_etal:2011} we model
the SH plate as a flat interface. In such a description,  we neglect an additional mechanism for a dissipation connected with the meniscus curvature.~\cite{sbragaglia.m:2007,harting.j:2008,ybert.c:2007} We emphasize however, that such an ideal situation is not unrealistic, and has been achieved in many recent experiments.~\cite{steinberger.a:2007,karatay.e:2013,haase.as:2013}

At the hydrodynamic level, the composite surface is modeled as spatially dependent
local slip boundary conditions:
\begin{equation}\left\{\begin{array}{ll}
\mathbf{u}_\tau-b(y)\dfrac{\partial \mathbf{u_ {\tau}}}{\partial z}=0,\\
u_z=0,
\end{array}\right.
\label{BCslip}\end{equation}
where $b$ is the slip length of the surface at point $(x,y)$, and  $\mathbf{u}_\tau=(u_x,u_y)$ is the tangential velocity of the fluid.
For smooth hydrophobic coating slip lengths are
 of the orders of tens of nm only,~\cite{vinogradova.oi:2009,vinogradova.oi:2003,charlaix.e:2005,joly.l:2006} so that for simplicity the boundary conditions at the solid areas are set as no-slip ($b=0$).

Note that prior work often assumed shear-free boundary condition over the gas sectors,~\cite{bocquet2007,priezjev.nv:2005,lauga.e:2003} so that the viscous
dissipation in the underlying gas phase has
been neglected. Here we use the partial slip boundary conditions. These are the consequence of the ``gas cushion model'', which takes into account that the
dissipation at the gas/liquid interface is dominated by the shearing of a
continuous gas layer~\cite{vinogradova.oi:1995a,andrienko.d:2003}

\begin{equation}\label{bgas}b(y)\simeq \dfrac{\mu}{\mu_{g}} e (y).\end{equation}
Here $\mu $ and $\mu_{g}$ are the liquid and the gas dynamic viscosities, respectively, and $e(y)$ the local thickness of the gas layer. Eq.(\ref{bgas}) represents an upper bound for a local slip at the gas area, which is attained in the limit of a small fraction of the solid phase. At a relatively large solid fraction, Eq.(\ref{bgas}) could overestimate the local slip.~\cite{ybert.c:2007} Beside that, net gas flux in the ``pockets'' of SH surfaces could be zero in case they are considered to have end walls,~\cite{maynes2007} which is also expected to increase the
surface friction and to decrease the slip. However, even in this extreme situation the local slip length profile, $b(y)$, necessarily follows the relief of the texture provided grooves are shallow enough (see Appendix~\ref{local}).

 A striped SH texture is fully determined by the fraction of the no-slip phase, $\phi_1=a/L$, and the local slip at the gas areas, which is proportional to the depth of the groove $b_{max} \simeq\dfrac{\mu}{\mu_{g}}e$. The trapezoidal texture provides an additional geometric parameter, the basement of the trapezoid $\alpha=c/L$, which is no longer equal to its top side, $\phi_1\leq\alpha\leq 1$ (see Fig. \ref{sketch_trap}).
Note that the striped SH surfaces represent a limiting case of trapezoidal textures with $\alpha=\phi_1$. The other limiting cases correspond to $\alpha=1$ (a vanishing distance between trapezoids) and $\phi_1=0$ (sawtooth-like profiles with a single solid-liquid contact point).

Using  Eq. (\ref{bgas}), we express the trapezoidal slip length profile in the form:
\begin{equation}b(t)=\left\{\begin{array}{ll}0,&|t|\leq \phi_1/2,\\
\dfrac{(2|t|-\phi_1)H\beta}{\alpha-\phi_1},&\dfrac{\phi_1}{2}\leq|t|\leq\dfrac{\alpha}{2}, \\
H\beta,&\dfrac{\alpha}{2}\leq|t|\leq\dfrac{1}{2}.
\end{array}\right.\label{btrap}\end{equation}
Here $t=\left[\dfrac{y}{L}\right]-\dfrac{1}{2}$ is a dimensionless variable defined on the periodic cell (brackets denote the fractional part of the number) and $\beta=\dfrac{b_{max}}{H}$ is the maximal slip length scaled by the channel thickness.

\section{Permeability of a thin superhydrophobic channel. \label{SecPerm}}

Here we provide some general theoretical results to guide the optimization of hydrodynamic phenomena in channels with one SH wall. Such a SH channel is characterized by three different length scales: the channel thickness $H$, the texture period $L$, and some macroscopic length $L_\infty\gg L$. Our analysis is based on the lubrication (or thin-channel) limit, where the texture varies over a scale $L \gg H$.  In this limit the pressure is uniform across the channel, and the flow profile $\mathbf{u}(x,y,z)$ is locally parabolic in $z$:
\begin{equation}\begin{array}{ll} \mathbf{u}(x,y,z)=-\dfrac{\nabla P(x,y)}{2\mu}\left[z+c_0(y)\right]\left(H-z\right),\;\end{array}\label{parab}\end{equation}
where the coefficient $c_0(y)$ is found from Eq. (\ref{BCslip}):
\begin{equation}
c_0(y)=\dfrac{b(y)H}{H+b(y)}.
\end{equation}
The depth-averaged velocity $\mathbf U(x,y)$ is then proportional to the local pressure gradient:
$$\mathbf{U}(x,y)=-k(y)\dfrac{\nabla P}{\mu},
$$
where $k(y)$ is the local permeability of the channel, related to the  slip length, $b(y)$, as
\begin{equation}\begin{array}{ll}
k(y)=\dfrac{H^2}{12}\dfrac{H+4b(y)}{H+b(y)}.\end{array}\label{k}\end{equation}

At the macroscopic scale,  $L_{\infty} \gg L$, local details of the flow become indiscernible, and the channel appears to be uniform and anisotropic. As usual for the lubrication approach, the relation between the macroscale (averaged over the period $L$) velocity $\mean\Ub$ and the macroscale pressure gradient $\mean{\nablab P}$
may be written as a Darcy law
 \[
\mean{\Ub} = - \frac{\kb_{\rm eff}}{\mu} \cdot \mean{\nablab P}.
\]
Here \[ \kb_{\rm eff} =\dfrac{H^2}{12} \left(\begin{array}{ll}k^*_\|& 0\\ 0 & k^*_\perp\end{array}\right),\]
is the  effective permeability tensor of the SH channel,~\cite{feuillebois.f:2009,schmieschek.s:2012}
 $k^*_\|$ and $k^*_\perp$ are the corrections to permeability of the hydrophilic channel of the same thickness, $H^2/12$, in the principal, i.e. longitudinal ($x$) and transverse  ($y$), directions (without transverse flow). These corrections can be found by solving two independent problems on a microlevel, for the pressure gradient directed along $x$- and $y$- axes,  and by averaging the resulting fields:  \begin{equation}\mean{ \mathbf U}=\dfrac{1}{L}\int_0^L \mathbf U(y)dy,\;\mean{\mathbf \nabla P}=\dfrac{1}{L}\int_0^L \mathbf \nabla p(y)dy,\label{micromacro}\end{equation}
where $\mathbf U(y), \nabla p(y)$ are solutions of Stokes equations at microlevel, which take into account all the details of the SH texture.

For the flow in the longitudinal direction the pressure gradient is uniform,  $\nabla_x p =\mean{\mathbf \nabla_x P}$. In the transverse configuration the total flow and thus the depth-averaged velocity is uniform, $U_y=\mean{U_y}$. This yields the following equations for the corresponding effective permeabilities~\cite{ajdari2002}
\begin{equation}
k_\|=\dfrac{1}{L}\int\limits_{0}^{L}k(y)dy,\hspace{1cm}k_\perp=\dfrac{1}{L}\left[\int\limits_0^L\frac{dy}{k(y)}\right]^{-1}.\label{kthin}
\end{equation}

For a thin channel with a striped SH wall the resulting formulae correspond to familiar limits
of resistors in parallel or in series.~\cite{feuillebois.f:2009} For the longitudinal configuration we get
\begin{equation}\label{upper}
 k_{U}=\phi_1k_1+\phi_2k_2,
\end{equation}
and for the transverse configuration
\begin{equation}\label{lower}
 1/k_{L}=\phi_1/k_1+\phi_2/k_2,
\end{equation}
where $\phi_2=1-\phi_1$ is the fraction of liquid surface in contact with gas and $k_{1,2}$ are local permeabilities of the corresponding parts of the channel, calculated using Eq. (\ref{k}). Values of $k_U$ and $k_L$ provide the upper and lower bounds for permeability for two-component texture with a piecewise constant slip $\beta$ and fixed fractions $\phi_1$, $\phi_2$.~\cite{feuillebois.f:2009}
Taking into account Eq. (\ref{k}) and applying boundary conditions one can obtain known formulae for the corrections to permeability in case of a SH channel with a striped wall:
\begin{equation}\begin{array}{ll}k^*_{U}(\beta,\phi_1)=\phi_1+\phi_2\dfrac{1+4\beta}{1+\beta},\\
k^*_{L}(\beta,\phi_1)=\left(\phi_1+\phi_2\dfrac{1+\beta}{1+4\beta}\right)^{-1},\end{array}\label{kstripes}\end{equation}
where $k_{U,L}= k^*_{U,L}H^2/12$.  Note that the bounds are fairly close when $\beta$ is very large, so in this situation the theory~\cite{feuillebois.f:2009} provides a good sense of the possible effective slip of any texture with a piecewise constant slip based only on the area fractions of gas and solid sectors, and on the local slip
length.

\section{Flow properties of a channel with a trapezoidal texture \label{SecProp}}

In this section, we consider a channel with one SH wall with the local slip length given by Eq. (\ref{btrap}).

In a lubrication limit, the permeability of such a channel can be calculated by using  Eq. (\ref{k}) and Eq. (\ref{kthin}):
\begin{equation}
\begin{array}{ll}k^*_\|=k^*_{U}-s(\beta)(\alpha-\phi_1)<k^*_{U},\\
k^*_\perp=\left[\dfrac{1}{k_L^{*}}-\dfrac{s(4\beta)}{4}\left(\alpha-\phi_1\right)\right]^{-1}<k^*_L,
\end{array}\label{ktrap}\end{equation}
 $$s=3\left[\dfrac{\ln\left(1+\beta\right)}{\beta}-\dfrac{1}{1+\beta}\right]>0.$$

Note that the difference between the permeabilities for the trapezoidal and striped channels with the same $\phi_1$ decays slowly with the increase in $\beta$ since $s(\beta)=O(\ln\beta/\beta)\to0$ as  $\beta\to\infty$. This means that for a very large $\beta$ there is almost no difference in permeability between striped and trapezoidal channels, provided $\phi_1$ is the same. However, the permeability of  channels with striped and trapezoidal walls will be different at finite $\beta$. Below we explore this in more detail.

\begin{figure}[h]
\includegraphics[width=0.46\textwidth]{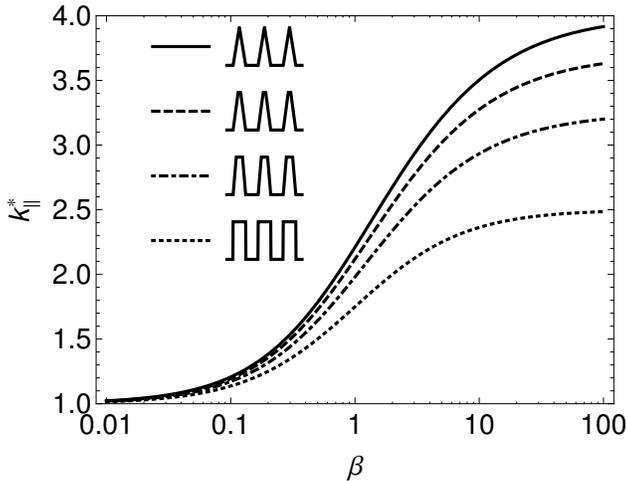}
\caption{ \it Corrections to permeability in the direction parallel to the texture for a trapezoidal wall with a fixed base width $\alpha=0.5$ and (from top to bottom) $\phi_1=0$, $0.1$, $0.25$ and $0.5$. }
\label{fig_k}
\end{figure}

  Fig. \ref{fig_k} shows the dependence $k^*_\|(\beta)$, which characterizes the maximal possible drag reduction in the channel, at a fixed value of the base width ($\alpha=0.5$) for different texture geometries, which obviously requires different $\phi_1$. (The results for a transverse configuration are qualitatively similar, with only
smaller values of $k^*_\perp$, this why we
do not show them here.)  We  vary $\phi_1$ in the interval from $0$ to $0.5$, so that we change the local slip profile from the triangular (saw-tooth texture) to the stripes. Regular trapezoidal SH
textures (with a finite solid area) then correspond to intermediate values of $\phi_1$. It can
be seen that qualitatively, results for all channels are similar to  a striped one. Namely, the correction to longitudinal permeability increases with $\beta$ and saturates at $\beta\to\infty$. However, it turns
out that the texture geometry (i.e. a decrease in $\phi_1$) dramatically enhances $k^*_\|$. Its maximal value can be evaluated with Eq. (\ref{ktrap}) and reads

\begin{equation}\label{beta_inf}
 k^*_\|(\beta\to\infty,\phi_1)=4-3\phi_1.
\end{equation}

\begin{figure}[h]
\includegraphics[width=0.46\textwidth]{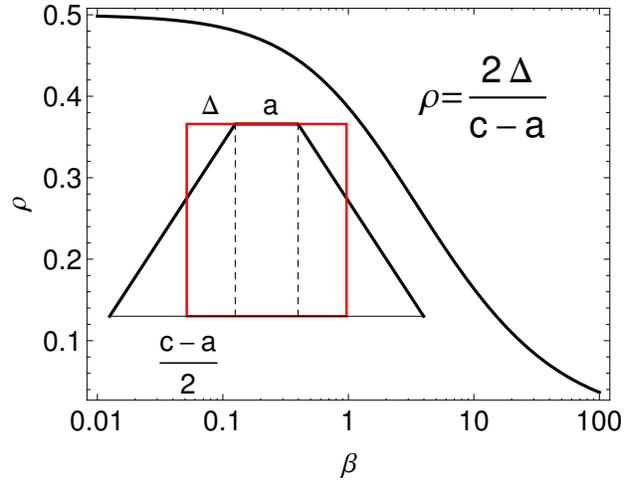}
\caption{ \it The function $\rho$ versus $\beta$. Inset illustrates the striped  (red) and the trapezoidal (black) local height profiles, leading to the same longitudinal permeability of the channel at a given $\beta$.    }
\label{fig_equiv}
\end{figure}

A similarity of the curves plotted in Fig. \ref{fig_k} suggests that we can establish a certain equivalence between trapezoidal and striped channels at a given $\beta$, since
 the correction to permeability for any trapezoidal texture will be the same as for a striped pattern, but with some different, ``apparent'' fraction of liquid/solid contact $\phi^*_1$. By using  Eq. (\ref{ktrap}) we can express this ``apparent'' fraction as:
\begin{equation}\phi_1^*=\phi_1+\rho(\beta)(\alpha-\phi_1),\label{equiv}
\end{equation}
\begin{equation}\label{rho}
  \rho=\dfrac{(1+\beta)\ln{(1+\beta)}}{\beta^2}-\dfrac{1}{\beta}.
\end{equation}

In particular, a channel with a partially slipping saw-tooth texture with $\phi_1=0$ and base $\alpha$ would be equivalent, from the point of view of a longitudinal permeability, to a striped channel with $\phi_1^*=\rho(\beta)\alpha$.

We emphasize that the analysis of function $\rho(\beta)$ allows one to interpret better the above results.  Indeed, one can easily demonstrate that $\rho(\beta)<1/2$ for any $\beta>0$. Therefore, the ``apparent'' fraction of no-slip surface, $\phi_1^*$, satisfies the following inequality (see Fig. \ref{fig_equiv}):
$$\phi_1^*<\dfrac{\phi_1+\alpha}{2}=\dfrac{a+c}{2L}.$$
This means that for equal average slip lengths $\mean{b}$ a trapezoidal pattern provides a larger $k^*_\|$. Beside that, $\rho(\beta)$ vanishes at large $\beta$, and $\phi_1^*\simeq\phi_1$, which clarifies the absence of a discernible difference in permeability between striped and trapezoidal channels at large local slip.

\begin{figure}[h]
\includegraphics[width=0.46\textwidth]{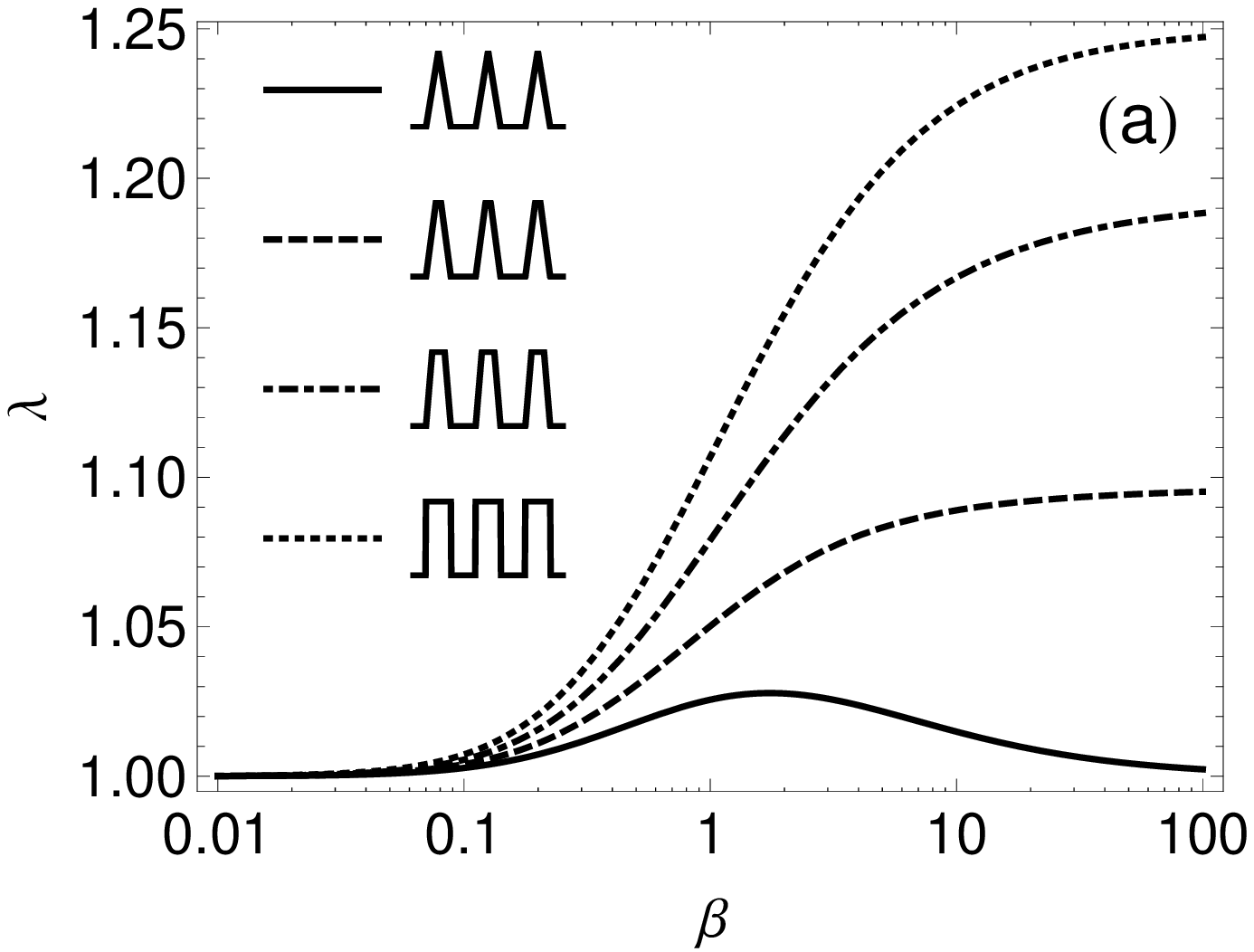}
\includegraphics[width=0.46\textwidth]{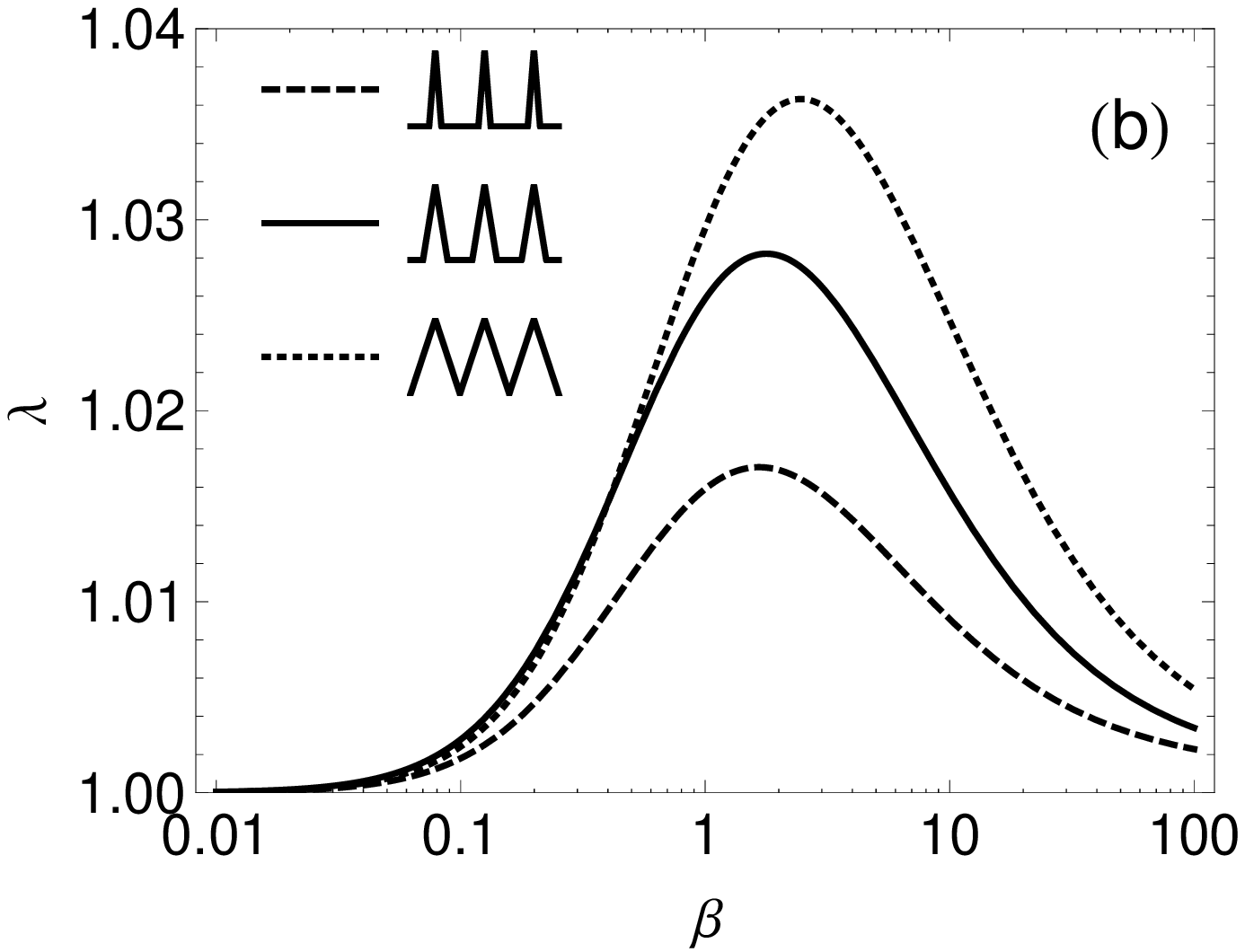}
\caption{ \it Anisotropy parameter for $(a)$ the same patterns as in Fig. \ref{fig_k}. $(b)$ saw-tooth textures with $\phi_1=0$ and (from top to bottom) $\alpha = 0.25,$ $0.5,$ $1$.}
\label{fig_lam}
\end{figure}

We remind that a pressure gradient in the eigendirections cannot produce any transverse flow. In other words, flow is aligned with forcing. In all other situations the direction
of the velocity $\mean{\Ub}$ is at some angle with respect to $\nabla P$ (see Fig. \ref{sketch_trap}). The maximum angle $\theta$ between  $\mean{\Ub}$ and  $\nabla P$ provides the largest transverse flow. The task of the optimization of its magnitude can be transformed to the maximization of a so-called anisotropy parameter
\begin{equation}\lambda=\sqrt{\dfrac{k^*_\|}{k^*_\perp}}\geq 1,\label{def_lam}\end{equation}
and then $\theta=\arctan(\lambda)$.~\cite{feuillebois.f:2010b}
The anisotropy parameter $\lambda$ for channels with a trapezoidal textured wall can be calculated by using Eqs. (\ref{ktrap}) and (\ref{def_lam}).

We first investigate the effect of $\alpha$ on the anisotropy parameter $\lambda$ for the same textures as in Fig. \ref{fig_k}. The results are shown in Fig.~\ref{fig_lam} (a).  We see that for all textures with large local slip $\lambda$ approaches its limit, which can be evaluated as
\begin{equation}\label{lambda_inf}
   \lambda(\beta\to\infty)=\dfrac{1}{2}\sqrt{4+9\phi_1-9\phi_1^2}.
\end{equation}
We have shown that the permeability of a thin SH channel with a trapezoidal wall is maximized by reducing the area fraction of solid, in contact
with the liquid (see Fig. \ref{fig_k}). In contrast, Fig.~\ref{fig_lam} (a) and Eq.(\ref{lambda_inf}) suggest that transverse flow in such channels
enhances for textures with a larger solid fraction. Thus, the transverse flow in thin channels is
maximized by stripes with a rather large solid fraction, $\phi_1=0.5$, and very large $\beta$. However, $\lambda$ becomes very small and even vanishes at large $\beta$ for a channel with a saw-tooth texture, $\phi_1=0$. For this texture the $\lambda$ reaches its maximum value at some $\beta=O(1)$. In contrast, for regular trapezoids (with finite $\phi_1$) $\lambda$ increases monotonously with $\beta$, which is qualitatively similar to a striped channel.

Next we examine the effect of varying $\alpha$ at fixed $\phi_1=0$.  Fig.~\ref{fig_lam} (b)
shows $\lambda$ as a function of $\beta$. The three
data sets correspond to different $\alpha$ and again show the maximum at some $\beta=O(1)$ and a decay to zero at large $\beta$.

\begin{figure}[h]
\includegraphics[width=0.46\textwidth]{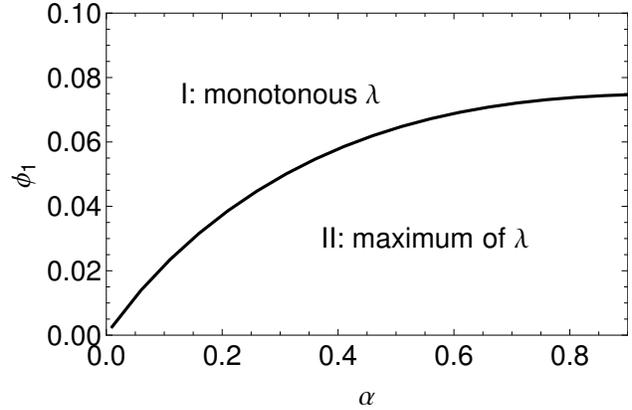}
\caption{ \it The boundary between two types of dependence of anisotropy parameter $\lambda(\beta)$.
\label{fig_diag}}
\end{figure}

The boundary between the two types of behavior of $\lambda$ can be found by using the conditions:
\begin{equation}\label{maximum}
  \begin{array}{ll}\dfrac{\partial\lambda(\beta,\phi_1,\alpha)}{\partial \beta}=0,&\dfrac{\partial^2\lambda(\beta,\phi_1,\alpha)}{\partial^2 \beta}=0\end{array}.
\end{equation}
Eqs. (\ref{maximum}) are solved numerically. The results are shown in Fig. \ref{fig_diag} and clearly demonstrate that a transition between two types of behavior of $\lambda$ happens at small $\phi_1$ and depends on a parameter $\alpha$. The upper region in Fig. \ref{fig_diag} corresponds to trapezoidal SH channels, where $\lambda$ increases with the maximal local slip, similarly to a classical striped channel. The lower region  corresponds to channels which show maximal transverse flow at some optimal finite $\beta$, similarly to a channel with the saw-tooth slip at the wall.

\begin{figure}[h]
\includegraphics[width=0.48\textwidth]{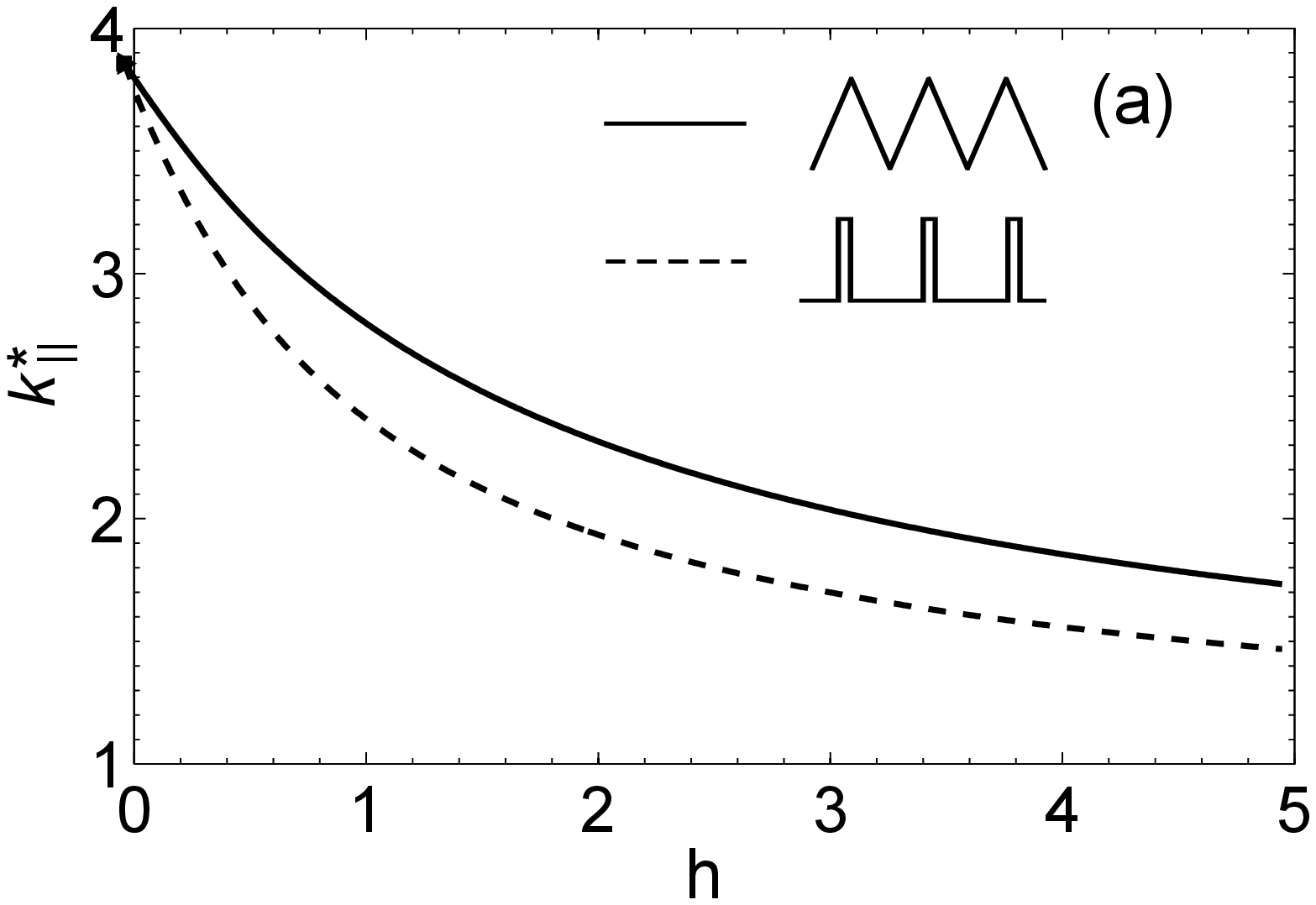}
\includegraphics[width=0.48\textwidth]{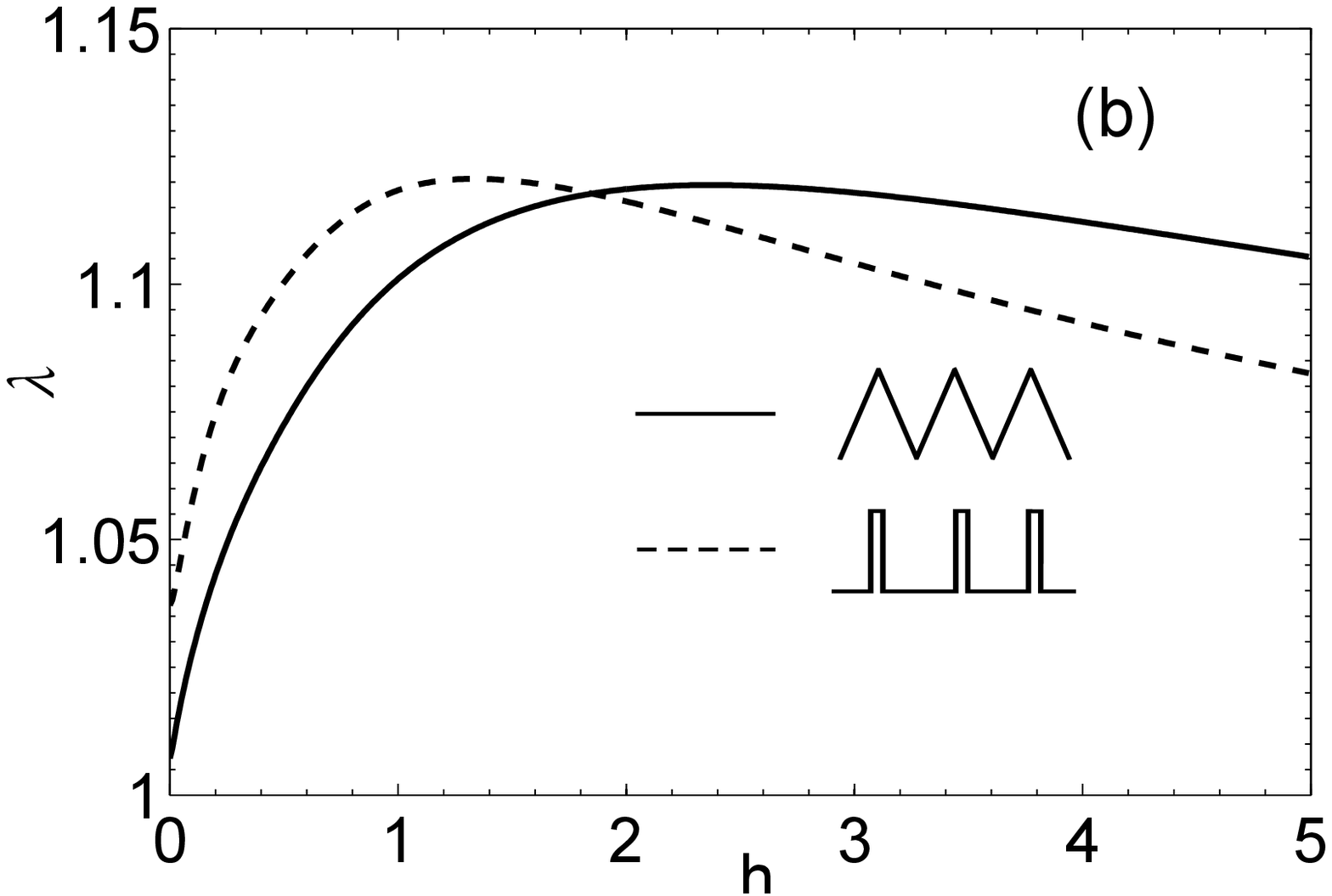}
\caption{ \it  Correction to a longitudinal permeability (a) and  the anisotropy parameter (b) as a function of $h=H/L$ calculated at fixed $\beta=50$. Solid curve shows the results for a channel with a saw-tooth texture ($\phi_1=0$, $\alpha=1$). Dashed curve presents results for an equivalent striped channel ($\phi_1^*\simeq 0.06$). }\label{FigH}
\end{figure}

The integral formulas for permeability (Eq. (\ref{k})) used above have been obtained in the limit of a thin channel. To estimate the range of validity of this approximation we solve the Stokes equations in the channel numerically and compare the calculated longitudinal permeability and anisotropy parameter with those obtained by using Eq. (\ref{ktrap}). Our numerical approach is described in Appendix B.

To test the analytical predictions we have chosen two ``extreme'' textures, which are equivalent in terms of $k^*_\|$ in a thin channel limit: a sawtooth texture with $\phi_1=0$, $\alpha=1$, $\beta=50$, and stripes with an equivalent $\phi^*_1\simeq0.06$, calculated using Eq. (\ref{equiv}). Due to a vanishing and low $\phi_1$ these patterns do not fully satisfy the assumptions of our thin channel theory, so that in this case the largest possible deviations from the analytical results are expected. In our numerical calculations we keep $\beta$ constant and vary $h=H/L$. This corresponds to varying texture period $L$ at fixed values of the groove depth $e$ and channel width $H$. Note that we can now relax the assumption of a thin channel to see if our analytical theory could be used outside the area of its formal applicability.

Fig.\ref{FigH} $(a)$ includes curves for the maximum permeability calculated for channels with the saw-tooth and equivalent striped textured wall. We remark that although the saw-tooth texture gives slightly greater permeability than the striped one, the deviations between them are small. This confirms our conclusion on the hydrodynamic equivalence of these two channels.  Both curves decay with $h$, which suggests that our analysis based on the lubrication theory overestimates $k^*_\|$.  However, at small $h$ results approach the predictions obtained within the lubrication  theory, so that it can be safely used even at the ``extreme'' situation of a vanishing solid area. At $h=O(1)$ the deviations from the theoretical predictions are significant, so that the thin-channel analysis cannot be applied.  We have performed similar calculation for different values of $\beta$, and have found similar behavior of $k_\|^*(h)$. This is why we do not show them here.

Fig.\ref{FigH} $(b)$ shows the anisotropy parameter for the same two channels. We see that the difference between channels with the saw-tooth and equivalent striped textured wall is again very small, so that both channels are expected to generate a comparable transverse flow. It can be seen that a thin channel approach underestimates $\lambda$, but could serve as its rough estimate when $h$ is relatively small even for channels with $\phi_1=0$. Finally, we note that at $h=O(1)$ the transverse flow is maximal, which suggests that channels of such a thickness may be promising for such microfluidic applications as separation or mixing. This subject however is outside of the scope of our paper and will be reported elsewhere.

\section{Conclusions \label{SecConc}}
We have studied hydrodynamic properties of a thin channel with one  wall decorated by a SH trapezoidal texture. General explicit expressions for the permeability tensor of a thin channel with an arbitrary texture have been formulated. These expressions, being applied for a SH trapezoidal channel, allowed us to calculate several important hydrodynamic properties. We have demonstrated that the trapezoidal textured wall of the channel can induce a rich and complex hydrodynamic behavior compared to expected for a classical striped channel, and can be advantageous in many situation. Furthermore, our theory gives analytical guidance as to how choose the parameters of the trapezoidal texture,
in order to optimize the forward and transverse flows. Our theoretical predictions have been compared with results
of numerical calculations, which have been also performed for thicker channels.
Our results are directly relevant for drag reduction and passive mixing in thin channels, as well as other microfluidic applications.

\section*{Acknowledgements}
This research was supported by the RAS through its priority program
``Assembly and Investigation of Macromolecular Structures of New
Generations'', and by the DFG through SFB 985.

\appendix
\section{Local slip conditions on the gas-liquid
interface for shallow gas cavity}\label{local}

Here we formulate the local slip conditions at the gas areas of a SH surface. We assume that the depth of the gas cavity is small, $\max \left[
e\left( y\right) \right] \ll L,$ i.e. the lubrication (thin-channel) limit can
be applied to the gas flow as well. Then the velocity profile $\mathbf{u}%
_{g}(x,y,z)$ is again locally parabolic in $z$ (cf. Eq.~(\ref{parab})):
\begin{equation}
\mathbf{u}_{g}(x,y,z)=-\dfrac{\mathbf{\nabla }P_{g}\left( x,y\right) }{2\mu
_{g}}\left[ z+c_{1}(y)\right] \left[ z+c_{2}(y)\right] .\;
\end{equation}%

\begin{figure}[h]
\includegraphics[width=0.46\textwidth]{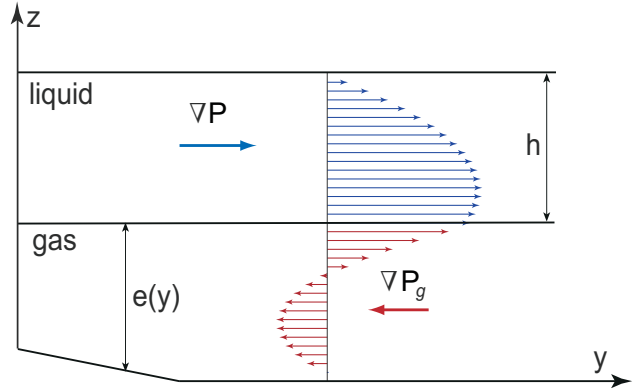}
\caption{ \it Schematic representation of a flow in a closed gas cavity with slowly varying depth. }\label{fig_gasflow}
\end{figure}

The velocity satisfies the no-slip condition at the bottom wall of the
cavity, $z=-e$, so we should require $c_{1}(y)=e(y).$ The gas cavity is
assumed to be closed in all horizontal directions. This
means that the pressure gradient in the gas $\mathbf{\nabla }P_{g}$ is such
 that the net flux of the gas flow is zero at any cross section (see Fig.~\ref{fig_gasflow}):%
\begin{equation}
\int_{-e}^{0}\left[ z+c_{1}(y)\right] \left[ z+c_{2}(y)\right] dz=0.
\label{flux}
\end{equation}%
One readily has from Eq.~(\ref{flux}) $$c_{2}(y)=e(y)/3.$$

The conditions at the gas-liquid interface are the continuity of the
velocity and the shear stress:
\begin{equation}
\begin{array}{l}
z=0:\quad u=u_{g}=-\dfrac{\nabla P_{g}e^{2}}{6\mu _{g}},\; \\
\qquad \mu \dfrac{\partial u}{\partial z}=\mu _{g}\dfrac{\partial u_{g}}{%
\partial z}=-\dfrac{2\nabla P_{g}e}{3}.%
\end{array}
\label{interface}
\end{equation}

Excluding $\nabla P_{g}$ in (\ref{interface}) we obtain
\begin{equation*}
z=0:\quad u=\frac{\mu e}{4\mu _{g}}\dfrac{\partial u}{\partial z},
\end{equation*}%
which is equivalent to the slip condition for the liquid flow with the local
slip length%
\begin{equation*}
b(y)=\frac{\mu e(y)}{4\mu _{g}}.
\end{equation*}%
$\allowbreak $
Thus the profile of local slip length follows the relief of shallow texture, but with the factor four times less than the ``gas cushion model'', Eq. (\ref{bgas}).

\section{Numerical method for calculating permeabilities in a channel of arbitrary gap}\label{numeric}
The numerical method is based on expansion of the solution into Fourier series and solving a linear system to satisfy boundary conditions Eq. (\ref{btrap}) on a spatial grid over $y\in[0,1/2]$. A similar method has been used in \cite{asmolov_etal:2013} to find the effective slip length for a wall-bounded linear shear flow.  The only difference here is that we keep both the growing and the decaying terms in $z$ and require them to annihilate at $z=h$. Also, we have to solve the problem for the transverse flow explicitly, because the relationship between longitudinal and transverse effective slip lengths derived in \cite{asmolov:2012} is not valid for a channel of a finite thickness.

We seek a correction to the Poiseuille flow driven by a fixed pressure gradient $\nabla P$:%
\begin{equation}\mathbf{u}=-\dfrac{H^2|\nabla P|}{2\mu}(\mathbf{v}_p+\mathbf{v}),\label{corr}\end{equation}
where $\mathbf v_p=z(h-z)/h^2\mathbf e_p$ is the undisturbed flow in non-dimensional variables ($z=Z/L$ and $\mathbf e_p$ is the direction of the pressure gradient) and $$\mathbf{v}=\left( v_x,v_y,v_z\right),$$ is the perturbation of the flow, caused by the presence of the texture.

The perturbation is zero at the upper wall (no-slip conditions) and satisfies the dimensionless Stokes equations,%
\begin{gather}
\mathbf{\nabla }\cdot \mathbf{v}=0,  \label{Se} \\
\mathbf{\nabla }p-\Delta \mathbf{v}=\mathbf{0},  \notag
\end{gather}%
where $p$ is pressure perturbation.
Our trapezoidal texture is periodic and symmetric in $y$: $b(y)=b(y+1)$, $b(1-y)=b(y)$. Therefore, the  solution to (\ref{Se}) for $\mathbf{v}$ and $p$ can be
represented in form of a Fourier series. To find the permeability tensor it is sufficient to consider two special cases: the pressure gradient directed parallel ($x$-direction) and perpendicular ($y$-direction) to the texture.

\subsection{Longitudinal configuration}

When the pressure gradient is directed along the texture, the
perturbation of the velocity has the only one component, $\mathbf{v}=\left( v_x,0,0\right)$, and due to the symmetry of $b(y)$ can be sought in terms of a cosine series:
$$v_x=\sum_{n=0 }^{\infty }v^\ast_x\left(
z,n\right) \cos \left(k_{n}y\right),$$
where $k_n=2\pi n$.

The Stokes equations Eq. (\ref{Se}) are then reduced
to a single ordinary differential equation (ODE):%
\begin{equation}
\Delta ^{\ast }v^{\ast }_x=0,  \label{1.6}
\end{equation}%
where $\Delta^*=\dfrac{d^2}{dz^2}-k_n^2$.
The zero-mode solution is
$$v^{\ast }_x(z,0)=c_0+c_1 z.$$
The solution for non-zero modes has the form
\begin{equation}
v^{\ast }_x(z,n)=c_{+}\left( n\right)\exp \left( k_{n} z \right)+c_{-}\left( n\right) \exp \left( - k_{n}
z\right).  \label{1x}
\end{equation}%
The no-slip boundary condition at the upper wall ($v^\ast_x=0$) provides a relation between the $c_{-}(n)$ and $c_{+}(n)$:
$$c_1=-c_0/h,\;c_{+}(n)=-\alpha_n c_{-}(n),$$
where $\alpha_n=\exp(-2k_nh)$ is the coefficient describing the influence  of the upper wall: for infinitely thick channel $\alpha_n=0$.

Now we can express the solution and its derivatives in terms of $c_0$ and $c_-(n)$ only. To obtain a linear system for these coefficients we truncate the series to $N$ terms and satisfy the partial slip boundary conditions Eq. (\ref{btrap}) in $N$ points distributed over half of the period, $y_j\in [0;1/2]$, $j=1..N$ (half of the period is taken since the slip length profile is symmetric, and the solution is sought in cosines only).
The resulting system of $N$ equations takes the following form:
\begin{equation}
\begin{array}{ll}
\displaystyle c_{0}\left[1+\dfrac{b(y_j)}{h}\right] +\sum_{n=1 }^{N-1}c_-(n)\left[f_n-b(y_j)g_n\right]\cos(k_ny_j)=\\=\dfrac{b\left( y_j\right)}{h},\;j=1..N,\end{array}
\label{SLAEpar}
\end{equation}%
where $f_n=(1-\alpha_n)$, $g_n=-(1+\alpha_n)k_n$.

The correction to channel permeability is obtained by averaging Eq.(\ref{corr}) over the channel gap and the texture period. Only the zero-mode term survives:
$$ k_{||}^*=1+3c_0.$$
The coefficient $c_0$ is found numerically by solving the linear system (\ref{SLAEpar}).

\subsection{Transverse configuration}

When the pressure gradient is directed across the texture, the velocity has two components, $\mathbf{v}=\left( 0,v_y,v_z\right)$, and the solution can be sought in terms of the Fourier coefficients as follows:
$$\begin{array}{ll}\displaystyle v_z=v_z^*(z,0)+\sum_{n=1}^{\infty}v_z^*(z,n)\sin(k_n y),\\\displaystyle v_y=v_y^*(z,0)+\sum_{n=1}^{\infty}v_y^*(z,n)\cos(k_n y),\\\displaystyle p=p^{\ast }(z,0)+\sum_{n=1}^{\infty}p^{\ast }(z,n)\sin(k_n y).\end{array}$$
%
The Stokes equations for transverse stripes can be written as:%
\begin{gather}
-k_{n}v_y^{\ast }+\frac{dv_z^{\ast }}{dz}=0,  \notag
\\
-k_{n}p^{\ast }-\Delta ^{\ast }v_y^{\ast }=0,  \label{2b} \\
\frac{dp^{\ast }}{dz}-\Delta ^{\ast }v_z^{\ast }=0.  \notag
\end{gather}%
for $n=1..N$. The solution for the zero-mode term is
$$v^{\ast }_z(z,0)=0,\; v^{\ast }_y(z,0)=d_0(1-z/h),\; p^{\ast }(z,0)= \mathrm{const}.  $$
By excluding $p^{\ast }\ $and $v^{\ast }_y,$ one can transform the equation for non-zero modes to a single ODE for the $z$-component:
\begin{gather}
\Delta ^{\ast 2}v^{\ast }_z=0.  \label{1.7}
\end{gather}%
A general solution of the fourth-order ODE (\ref{1.7})
can be written in the following form:%
\begin{equation*}
v^{\ast }_z=\left( c_{-}+d_{-}z\right) \exp \left( -
k_{n} z\right)+ \left( c_{+}+d_{+}z\right) \exp \left(
k_{n} z\right),
\end{equation*}%
which involves four unknown constants per mode.
The Fourier coefficients for the $y$-component can be obtained from Eq.~(\ref{2b}):
\begin{equation*}
v^{\ast }_y=\dfrac{1}{k_n}\dfrac{d v^*_z}{d z}.
\end{equation*}%
The no-slip and impermeability conditions read
$$v_z^*(0,n)=0,\;v_z^*(h,n)=0, \;v_y^*(h,n)=0.$$
They provide three conditions per mode and enable us to express all the unknown coefficients in terms of $d_{-}(n)$.
To satisfy the partial slip boundary conditions we have the following linear system:
\begin{equation}
\begin{array}{ll}\displaystyle d_{0}\left[1+\dfrac{b(y_j)}{h}\right] +\sum_{n=1 }^{N-1}d_-(n)\left[f_n-b(y_j)g_n\right]\cos(k_ny_j) =\\=\dfrac{b\left( y_j\right)}{h},\;j=1..N,\end{array}
\label{ls1}
\end{equation}%
where
$$f_n=\dfrac{2\alpha_n-\alpha_n^2-1+ \gamma_n^2\alpha_n}{k_n(\alpha_n+\gamma_n\alpha_n-1)},$$
$$g_n=-\dfrac{2(\alpha_n^2-1+2\gamma_n\alpha_n)}{(\alpha+\gamma_n\alpha_n-1)},$$
$$\gamma_n=2k_nh.$$
At large $h$ we have $p_n=1$ and $q_n=-2k_n$, in full agreement with the result for the infinite channel.\cite{asmolov:2012}
The linear system (\ref{ls1}) is solved numerically, and the correction to transverse permeability is calculated as
$$ k_{\perp}^*=1+3d_0.$$

\footnotesize{

\providecommand*{\mcitethebibliography}{\thebibliography}
\csname @ifundefined\endcsname{endmcitethebibliography}
{\let\endmcitethebibliography\endthebibliography}{}

\bibliographystyle{rsc} 

}
\end{document}